\documentclass{iopart}

\usepackage{iopams}
\usepackage{graphicx}  
\usepackage{subfigure}
\usepackage{sidecap}

\begin{document}

\title{Proximity effect in superconductor / conical magnet
  heterostructures}

\author{Daniel Fritsch and James F. Annett}

\address{H. H. Wills Physics Laboratory, School of Physics, University
  of Bristol, Bristol BS8 1TL, UK}

\ead{daniel.fritsch@bristol.ac.uk}

\begin{abstract}
The presence of a spin-flip potential at the interface between a
superconductor and a ferromagnetic metal allows for the generation of
equal-spin spin-triplet Cooper pairs. These Cooper pairs are
compatible with the exchange interaction within the ferromagnetic
region and hence allow for the long-range proximity effect through a
ferromagnet or half-metal. One suitable spin-flip potential is
provided by incorporating the conical magnet Holmium (Ho) into the
interface. The conical magnetic structure is characterised by an
opening angle $\alpha$ with respect to the crystal $c$-axis and a
turning (or pitch) angle $\beta$ measuring the rotation of
magnetisation with respect to the adjacent layers. Here, we present
results showing the influence of conical magnet interface layers with
varying $\alpha$ and $\beta$ on the efficiency of the generation of
equal-spin spin-triplet pairing. The results are obtained by
self-consistent solutions of the microscopic Bogoliubov$-$de Gennes
equations in the clean limit within a tight-binding model of the
heterostructure. In particular, the dependence of unequal-spin and
equal-spin spin-triplet pairing correlations on the conical magnetic
angles $\alpha$ and $\beta$ are discussed in detail.
\end{abstract}

\submitto{\JPCM}

\section{Introduction}
\label{Introduction}

The interface of a superconductor (SC) and a ferromagnetic metal (FM)
is a source of many intriguing
phenomena~\cite{Buzdin_RMP77_935,Bergeret_RMP77_1321,Tanaka_JPSJ81_011013,Eschrig_JLowTempPhys147_457}. Unlike
the case of the proximity effect between a SC and a normal metal,
there is oscillation of the decaying Cooper pair density in the
ferromagnetic region of the heterostructure. This oscillation occurs
because of the slightly different Fermi wave-vectors of the electrons
of different spin in the FM. In the SC/FM proximity effect these
oscillations occur for both the spin-singlet part of the pairing
correlation, and also for the opposite spin ($m = 0$ relative to the
FM magnetisation direction) part of the spin-triplet pairing
correlations. Both of these parts of the pairing correlations are
strongly suppressed by non-magnetic disorder in the FM and so lead to
only a short ranged proximity effect in SC/FM heterostructures. If,
however, the SC/FM interface allows for some kind of spin-flip
process, then in addition to these singlet and opposite-spin triplet
Cooper pair correlations it is also possible that equal-spin
spin-triplet pairing correlations ($m = \pm 1$) can be generated at
the interface. Such pairing correlations are compatible with the
ferromagnetic exchange interaction~\cite{Bergeret_PRL86_4096} and
hence do not oscillate or decay in the FM region. These equal-spin
spin-triplet pairing correlations thus allow for the long-range
proximity effect, which is observed in certain SC/FM
heterostructures. A review summarising the proximity effect in SC/FM
heterostructures is provided by Buzdin~\cite{Buzdin_RMP77_935},
whereas Bergeret \textit{et al.}~\cite{Bergeret_RMP77_1321} and Tanaka
\textit{et al.}~\cite{Tanaka_JPSJ81_011013} discuss in detail this new
type of ``odd-triplet'' superconductivity and Eschrig \textit{et
  al.}~\cite{Eschrig_JLowTempPhys147_457} provide detailed information
about the underlying symmetry relations.

Experimentally, several different approaches to provide a spin-flip
process at the interface have been proposed and proven to generate
long-ranged SC/FM proximity effects.  Here we specifically focus on
the case where the necessary spin-flip scattering potential is
provided by conical magnetic interface layers. Long-ranged proximity
effects have been observed in experiments on heterostructures
containing the conical ferromagnet Holmium (Ho), by Sosnin \textit{et
  al.}~\cite{Sosnin_PRL96_157002}, Hal\'{a}sz \textit{et
  al.}~\cite{Halasz_PRB79_224505,Halasz_PRB84_024517}, and Robinson
\textit{et al.}~\cite{Robinson_Science329_59}. Theoretically, Ho
containing heterostructures have been investigated by Alidoust
\textit{et al.}~\cite{Alidoust_PRB81_014512}, Wu \textit{et
  al.}~\cite{Wu_PRB86_184517} and in our previous
work~\cite{Fritsch_Arxiv2013}. These calculations confirm that the
conical magnet acts as a suitable spin-flip potential and generates
equal-spin spin-triplet pairing correlations which extend far into the
FM region, providing the microscopic basis for the long ranged
proximity effect in such heterostructures.

The aim of this paper is to investigate in greater detail a single
interface between an SC and a conical magnet (CM). The conical
magnetic structure of Ho is characterised by the opening angle $\alpha
= 80\,^{\circ}$ measuring the deviation of magnetisation from the
$c$-axis growth direction (assumed normal to the SC/FM interface) and
a turning (or pitch) angle $\beta = 30\,^{\circ}$ describing the
magnetisation rotation between adjacent Ho layers. All previous
theoretical investigations of SC/CM proximity effects focus on
heterostructures containing Ho, and so naturally the conical magnetic
angles $\alpha$ and $\beta$ have been kept fixed to their respective
Ho values. Consequently, an overall picture of the effectiveness of
equal-spin spin-triplet generation of Ho compared to other possible
conical magnetic interface materials has not emerged yet. Here we
present calculations showing the influence of conical magnetic
interface layers with varying $\alpha$ and $\beta$ on the efficiency
of equal-spin spin-triplet generation. We examine SC/CM interfaces in
which the conical magnet parameters $\alpha$ and $\beta$ are varied in
the ranges of $0\,^{\circ}$ to $180\,^{\circ}$. We have performed
calculations for the full range of $\alpha$ and $\beta$ angles with a
step size of $5\,^{\circ}$. For each of the parameter sets the
microscopic Bogoliubov$-$de Gennes (BdG) equations have been solved
self-consistently in the clean limit.

The paper is organised as follow. \Sref{Sec2} provides the theoretical
background, including the necessary description of the microscopic BdG
equations in \sref{Sec2_1}, and the method used to calculate the
spin-triplet pairing correlations in \sref{Sec2_2}. Starting with the
conical magnetic structure of Ho ($\alpha = 80\,^{\circ}$ and $\beta =
30\,^{\circ}$) we first show, in \sref{Sec3_1}, results for these
values of $\alpha$ and $\beta$ to introduce the key calculated
quantities and our basic notations. Keeping $\alpha$ fixed the
influence of varying $\beta$ is then discussed in
\sref{Sec3_2}. Finally \sref{Sec3_3} shows results obtained for
varying both $\alpha$ and $\beta$ simultaneously. A summary and
outlook are provided in the conclusion \sref{SummaryAndOutlook}.

\section{Theoretical Background}
\label{Sec2}

\subsection{Bogoliubov$-$de Gennes equations and heterostructure setup}
\label{Sec2_1}

The results presented below utilise self-consistent solutions of the
microscopic BdG equations in the clean limit. The method employed is
similar to our previous work on Ho containing
heterostructures~\cite{Fritsch_Arxiv2013}, so only the most relevant
formulas are repeated here. Generally, the spin-dependent BdG
equations can be written
as~\cite{Fritsch_Arxiv2013,Sipr_JPCM7_5239,Annett_book,KettersonSong_Superconductivity}
\begin{eqnarray}
  \fl \label{EqBdGGeneral} \left(
  \begin{array}{cccc}
    {\cal H}_{0} - h_{z} & -h_{x} + i h_{y} & \Delta_{\uparrow
      \uparrow} & \Delta_{\uparrow \downarrow} \\ - h_{x} - i h_{y} &
    {\cal H}_{0} + h_{z} & \Delta_{\downarrow \uparrow} &
    \Delta_{\downarrow \downarrow} \\ \Delta_{\uparrow \uparrow}^{*} &
    \Delta_{\downarrow \uparrow}^{*} & -{\cal H}_{0} + h_{z} & h_{x} +
    i h_{y} \\ \Delta_{\uparrow \downarrow}^{*} & \Delta_{\downarrow
      \downarrow}^{*} & h_{x} - i h_{y} & -{\cal H}_{0} - h_{z}
  \end{array}
  \right) \left(
  \begin{array}{c}
    u_{n\uparrow} \\ u_{n\downarrow} \\ v_{n\uparrow}
    \\ v_{n\downarrow}
  \end{array}
  \right) =\varepsilon_{n} \left(
  \begin{array}{c}
    u_{n\uparrow} \\ u_{n\downarrow} \\ v_{n\uparrow}
    \\ v_{n\downarrow}
  \end{array}
  \right) \,.
\end{eqnarray}
Here, $\varepsilon_{n}$ denotes the eigenvalues, and $u_{n\sigma}$ and
$v_{n\sigma}$ are quasiparticle and quasihole amplitudes for spin
$\sigma$, respectively. After some
simplifications~\cite{Fritsch_Arxiv2013,Sipr_JPCM7_5239,Covaci_PRB73_014503}
the tight-binding Hamiltonian ${\cal H}_{0}$ reads
\begin{equation}
  \label{EqTBHamiltonianLinear}
        {\cal H}_{0} = -t \sum_{n}{ \left( c_{n}^{\dagger}c_{n+1} +
          c_{n+1}^{\dagger}c_{n} \right) } + \sum_{n}{ \left(
          \varepsilon_{n} - \mu \right) c_{n}^{\dagger} c_{n} } \,,
\end{equation}
with $c_{n}^{\dagger}$ and $c_{n}$ being electronic creation and
destruction operators at multilayer index $n$, respectively. Choosing
the next-nearest neighbour hopping parameter $t = 1$ and the chemical
potential (Fermi energy) $\mu = 0$ sets the energy scales.

The vector components of the conical exchange field are written
as~\cite{Alidoust_PRB81_014512,Wu_PRB86_184517,Fritsch_Arxiv2013}
\begin{equation}
\label{EqHelicalMagnet}
  {\bf h} = h_{0} \left\{ \cos \alpha {\bf y} + \sin \alpha \left[
    \sin \left( \frac{\beta y}{a} \right) {\bf x} + \cos \left(
    \frac{\beta y}{a} \right) {\bf z} \right] \right\} \,,
\end{equation}
with $h_{0} = 0.1$ being the strength of the conical magnet's exchange
field and $a = 1$ being the lattice constant.

The general form of the pairing matrix in \eref{EqBdGGeneral} can be
rewritten according to Balian and
Werthamer~\cite{Balian_PhysRev131_1553,Sigrist_RMP63_239}
\begin{equation}
  \label{EqBalianWerthamer} \left( \begin{array}{cc}
    \Delta_{\uparrow \uparrow} & \Delta_{\uparrow \downarrow}
    \\ \Delta_{\downarrow \uparrow} & \Delta_{\downarrow
      \downarrow} \end{array} \right) = \left( \Delta + {\hat{\bf
      \sigma}} \cdot {\bf d}\right) i \hat{\sigma}_{2} =
  \left( \begin{array}{cc} -d_{x} + i d_{y} & \Delta + d_{z} \\ -
    \Delta + d_{z} & d_{x} + i d_{y} \end{array} \right) \,,
\end{equation}
with $\hat{\cdots}$ indicating a $2 \times 2$ matrix. Utilising the
Pauli matrices $\hat{\bf \sigma}$, the superconducting order parameter
is described by a singlet (scalar) part $\Delta$ and a triplet
(vector) part ${\bf d}$, respectively.

\begin{SCfigure}
  \centering
  \includegraphics[width=0.65\textwidth,clip]{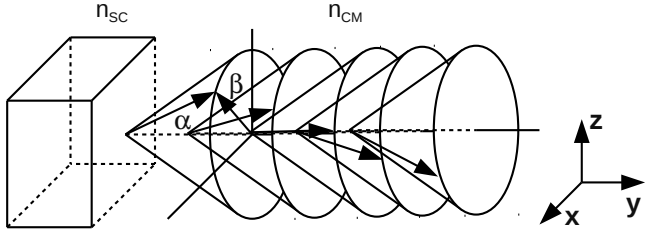}
  \fl \caption{\label{Fig1} Setup of heterostructure consisting of a
    spin-singlet $s$-wave superconductor ($n_{\rm SC} = 250$ layers),
    and a conical magnet ($n_{\rm CM} = 500$ layers). The opening and
    turning angles $\alpha$ and $\beta$ of the conical magnet are
    defined as shown. From \eref{EqHelicalMagnet} it follows that
    $\alpha$ is measured from $+y$ towards $+z$, whereas $\beta$ is
    measured from $+z$ towards $+x$ in the coordinate system
    indicated.}
\end{SCfigure}

In our model the pairing interaction is assumed to apply only for
$s$-wave singlet Cooper pairs, and so the pairing potential is
restricted to a scalar quantity $\Delta$. This fulfils the
self-consistency condition
\begin{equation}
  \label{EqDeltaSelfConsistency}
  \Delta({\bf r}) = \frac{g({\bf r})}{2} \sum_{n}{ \bigl(
    u_{n\uparrow}({\bf r})v_{n\downarrow}^{*}({\bf r})
    [1-f(\varepsilon_{n})] + u_{n\downarrow}({\bf
      r})v_{n\uparrow}^{*}({\bf r}) f(\varepsilon_{n}) \bigr)} \,,
\end{equation}
where we are summing only over positive eigenvalues $\varepsilon_{n}$,
and where $f(\varepsilon_{n})$ denotes the Fermi distribution function
evaluated as a step function for zero temperature. Setting up the
heterostructure as shown in \fref{Fig1} the effective superconducting
coupling parameter $g({\bf r})$ equals $1$ in the $n_{\rm SC} = 250$
layers of spin-singlet $s$-wave superconductor to the left of the
interface and vanishes elsewhere.
The conical exchange field, defined according to
\eref{EqHelicalMagnet}, is added to the $n_{\rm CM} = 500$ layers of
conical magnet to the right of the interface.

\subsection{(Triplet) Pairing correlations}
\label{Sec2_2}

The superconducting pairing correlation between spins $\alpha$ and
$\beta$ can generally be evaluated as on-site interaction for times $t
= \tau$ and $t' = 0$ as
\begin{equation}
  \label{EqPairingCorrelationGeneral}
  f_{\alpha \beta}({\bf r}, \tau, 0) = \frac{1}{2}\bigl<
  \hat{\Psi}_{\alpha}({\bf r},\tau) \hat{\Psi}_{\beta}({\bf r},0)
  \bigr> \,,
\end{equation}
with $\hat{\Psi}_{\sigma}({\bf r},\tau)$ being the many-body field
operator for spin $\sigma$ at time $\tau$. The time-dependence is
introduced through the Heisenberg equation of motion. Being local in
space the pairing correlation evaluated using
\eref{EqPairingCorrelationGeneral} leads to vanishing triplet
contribution for $\tau = 0$ in accordance with the Pauli
principle~\cite{Halterman_PRL99_127002}. However for finite times
$\tau$ nonvanishing contributions emerge, an example of odd-frequency
triplet pairing~\cite{Bergeret_RMP77_1321}. Substituting the field
operators valid for our setup and phase convention the spin-dependent
triplet pairing correlations are given by
\begin{equation}
  \label{EqTripletPairingCorrelations}
  \fl \eqalign { f_{0}(y, \tau) = \frac{1}{2} \bigl( f_{\uparrow
      \downarrow}(y, \tau) + f_{\downarrow \uparrow}(y, \tau) \bigr) =
    \frac{1}{2} \sum_{n} {\bigl( u_{n\uparrow}(y)
      v_{n\downarrow}^{*}(y) + u_{n\downarrow}(y) v_{n\uparrow}^{*}(y)
      \bigr) \zeta_{n}(\tau)} \cr f_{1}(y, \tau) = \frac{1}{2} \bigl(
    f_{\uparrow \uparrow}(y, \tau) - f_{\downarrow \downarrow}(y,
    \tau) \bigr) = \frac{1}{2} \sum_{n} {\bigl( u_{n\uparrow}(y)
      v_{n\uparrow}^{*}(y) - u_{n\downarrow}(y) v_{n\downarrow}^{*}(y)
      \bigr) \zeta_{n}(\tau)} }
\end{equation}
depending on position $y$ and time parameter $\tau$ (set to $\tau =
10$ in the present numerical work), and with $\zeta_{n}(\tau)$ given
by
\begin{equation}
  \label{EqTau}
  \zeta_{n}(\tau) = \cos (\varepsilon_{n} \tau) - i \sin
  (\varepsilon_{n} \tau) \bigl( 1 - 2 f(\varepsilon_{n}) \bigr) \,.
\end{equation}

The pairing amplitudes in \eref{EqTripletPairingCorrelations} are
defined relative to a single spin-quantisation direction, here $z$. In
the SC/FM proximity effect this has a natural choice as the FM spin
direction. However in the present calculations for a SC/CM interface
there is no such fixed reference axis, since the rotation of the CM
moment leads to a spatially varying local magnetisation
direction. Therefore it is helpful to represent the triplet pairing
correlations in a form which is independent of the spin-quantisation
axis chosen. Recalling \eref{EqBalianWerthamer} we define ${\hat
  \Delta}$ as the triplet pairing matrix for an ordinary spin-triplet
superconductor
\begin{equation}
  \label{EqBalianWerthamer2}
        {\hat \Delta} = \left( {\hat{\bf \sigma}} \cdot {\bf d}\right)
        i \hat{\sigma}_{2} = \left( \begin{array}{cc} -d_{x} + i d_{y}
          & d_{z} \\ d_{z} & d_{x} + i d_{y} \end{array} \right) \,,
\end{equation}
which obeys the following identity~\cite{Sigrist_RMP63_239}
\begin{equation}
  \label{EqDDdagger}
        {\hat \Delta}{\hat \Delta}^{\dagger} = |{\bf d}|^{2} {\hat
          \sigma}_{0} + i \left( {\bf d}\times{\bf d}^{*} \right)
        \cdot {\hat {\bf \sigma}} \,,
\end{equation}
where ${\hat \sigma}_{0}$ denotes the identity matrix. For bulk
triplet superconductors unitary pairing states are defined by the
condition $i{\bf d}\times{\bf d}^{*} = 0$ and for these states $2|{\bf
  d}|$ is the quasiparticle energy gap. For non-unitary triplet states
$i{\bf d}\times{\bf d}^{*} \neq 0$ and is a (real) vector in the
direction of the net Cooper pair spin, i.e., it is a measure of the
pair spin magnetic
moment~\cite{Sigrist_RMP63_239,Alidoust_PRB81_014512}. The quantities
$|{\bf d}|$ and $i{\bf d}\times{\bf d}^{*}$ are obviously also gauge
invariant, unlike the complex vector ${\bf d}$ itself.

In the present work there is no pairing interaction in the triplet
channel, \eref{EqDeltaSelfConsistency}, and so ${\bf d} =
0$. Therefore, instead of this we focus on the triplet pair
correlation function matrix which can be written similarly to the
matrix \eref{EqBalianWerthamer2}
as~\cite{Kawabata_JPSJ82_124702,Alidoust_PRB81_014512}
\begin{equation}
  \label{EqPairFunctionMatrix}
        {\hat f} = \left( {\hat{\bf \sigma}} \cdot {\bf f} \right) i
        {\hat \sigma}_{2} = \left( \begin{array}{cc} -f_{x} + i f_{y}
          & f_{z} \\ f_{z} & f_{x} + i f_{y} \end{array} \right) \,.
\end{equation}
(Note the additional factor $i$ in this definition compared
to~\cite{Kawabata_JPSJ82_124702} to be consistent with the definition
of ${\hat \Delta}$ in \eref{EqBalianWerthamer}.) Now the analogue to
\eref{EqDDdagger} reads
\begin{equation}
  \label{EqFFdagger}
        {\hat f}{\hat f}^{\dagger} = |{\bf f}|^{2} {\hat \sigma}_{0} +
        i \left( {\bf f}\times{\bf f}^{*} \right) \cdot {\hat {\bf
            \sigma}} \,.
\end{equation}
The two analogues to $|{\bf d}|$ and ${\bf d}\times{\bf d}^{*}$,
namely $|{\bf f}|$ and ${\bf f}\times{\bf f}^{*}$, can be expressed
conveniently in terms of the ${\bf f}$-vector components, depending on
the spin-dependent triplet pairing correlations as
\begin{equation}
  \label{EqDVector}
  \eqalign { f_{x} = \frac{1}{2} \left( -f_{\uparrow \uparrow} +
    f_{\downarrow \downarrow} \right) \cr f_{y} = - \frac{i}{2} \left(
    f_{\uparrow \uparrow} + f_{\downarrow \downarrow} \right) \cr
    f_{z} = \frac{1}{2} \left( f_{\uparrow \downarrow} + f_{\downarrow
      \uparrow} \right) } \,.
\end{equation}
The coordinate axis independent quantity, $|{\bf f}|^2$, can be
interpreted as the mean density of spin-triplet Cooper pairs, and the
quantity $i{\bf f}\times{\bf f}^{*}$ can be interpreted as a vector
whose magnitude and direction denote the net spin moment arising from
the non-unitary (equal-spin) part of the triplet Cooper pair
density. Of course the quantities $|{\bf f}|$ and $i{\bf f}\times{\bf
  f}^{*}$ are also gauge invariant under overall phase changes of the
superconducting order parameter.

\section{Results and Discussion}
\label{Sec3}

\subsection{Fixed $\alpha$, fixed $\beta$}
\label{Sec3_1}

The results presented in this section are for pairing correlations at
a SC/CM interface, obtained via a parameter scan of opening angle
$\alpha$ and turning (or pitch) angle $\beta$ in the ranges from
$0\,^{\circ}$ to $180\,^{\circ}$, calculated in steps of
$5\,^{\circ}$. According to the definition \eref{EqHelicalMagnet} of
$\alpha$, conical magnetic structures with $\alpha < 90\,^{\circ}$
($\alpha > 90\,^{\circ}$) have magnetic moment orientations pointing
away (towards) the superconducting side of the interface. $\alpha =
90\,^{\circ}$ denotes the special case of a helical magnet with
$\beta$, now known as the pitch angle. For the limiting cases with
$\beta = 0\,^{\circ}$ ($\beta = 180\,^{\circ}$) the magnetic moments
between adjacent layers are ordered ferromagnetically
(antiferromagnetically), respectively.

\begin{figure}
  \centering
  \includegraphics[width=0.85\textwidth,clip]{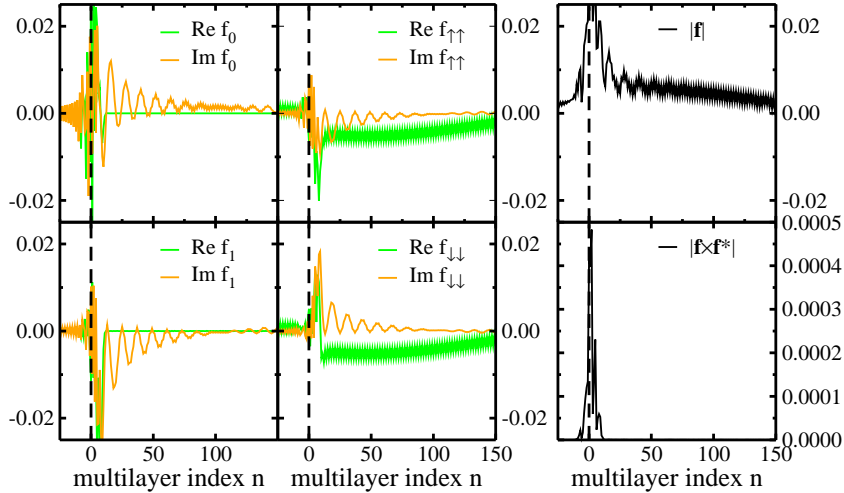}
  \caption{\label{Fig2} Real (green) and imaginary (orange) parts of
    the triplet pairing correlations $f_{0}$ (upper left panel) and
    $f_{1}$ (lower left panel) according to
    \eref{EqTripletPairingCorrelations} for $\alpha = 80\,^{\circ}$
    and $\beta = 30\,^{\circ}$. For better analysis the upper (lower)
    middle panels depict the real (green) and imaginary (orange) parts
    of $f_{\uparrow \uparrow}$ ($f_{\downarrow \downarrow}$) as
    components of $f_{1}$ shown in the lower left panel. Upper and
    lower right panels show the magnitude of the ${\bf f}$-vector and
    the magnitude of ${\bf f}\times{\bf f}^{*}$ as introduced in
    \eref{EqFFdagger}, respectively. All data is shown depending on
    the multilayer index $n$, with the interface positioned at zero
    (dashed line).}
\end{figure}

The fundamental pairing amplitudes which we calculate are shown in
\fref{Fig2}, which shows results for a SC/CM interface containing Ho
as the conical magnet, i.e. fixing $\alpha = 80\,^{\circ}$ and $\beta
= 30\,^{\circ}$.
The upper (lower) left panels show the real (green) and imaginary
(orange) parts of the equal-spin (unequal-spin) spin-triplet
correlations $f_{0}$ ($f_{1}$) as defined according to
\eref{EqTripletPairingCorrelations}. The interface is indicated by the
dashed line at layer $n = 0$, with the superconductor on the left
(layer $n < 0$) and the CM on the right ($n \geq 0$). The upper
(lower) middle panels depict the real (green) and imaginary (orange)
parts of $f_{\uparrow \uparrow}$ ($f_{\downarrow \downarrow}$) for the
same system. According to \eref{EqTripletPairingCorrelations}, these
pairing correlations are the two contributions to $f_{1}$. Finally,
the upper and lower right panels of \fref{Fig2} depict the magnitude
of the ${\bf f}$-vector and the magnitude of ${\bf f}\times{\bf
  f}^{*}$ as introduced in \eref{EqFFdagger}, respectively.

Starting with $f_{0}$ and $f_{1}$, shown in the left panels of
\fref{Fig2}, one notices only slight oscillations of the real parts
around the interface layer, whereas the oscillations of the imaginary
parts proceed well into the conical magnet layer. It has to be noted
that the faster oscillations visible in the imaginary part are related
to the chosen turning angle $\beta$, i.e., the layer index difference
between adjacent maxima correspond to one full turn of the conical
magnetic structure. This is superimposed on a slower oscillation
responsible for some pronounced intensity decay in the imaginary parts
of $f_{0}$ and $f_{1}$ as one goes further into the conical magnet
side of the interface.

Looking now at the middle panels of \fref{Fig2}, depicting the real
and imaginary parts of $f_{\uparrow \uparrow}$ and $f_{\downarrow
  \downarrow}$, one notices a different behaviour for real and
imaginary parts. The real parts of these quantities differ only in the
interface region, and are of equal sign and size in parts of the
conical magnet away form the interface. Evaluating $f_{1}$ according
to \eref{EqTripletPairingCorrelations} leads to the vanishing
contributions away from the interface. The very fast oscillations
shown in the real parts away from the interface are an artefact of the
simple model used here and should vanish once the full ${\bf k}$
dependence (parallel to the interface layer) and / or impurity
scattering effects are taken into account, thereby allowing for
deviations from the clean limit. The imaginary parts of $f_{\uparrow
  \uparrow}$ and $f_{\downarrow \downarrow}$, however, are of equal
size but opposite sign leading to the nonvanishing contributions to
$f_{1}$ shown in the lower left panel of \fref{Fig2}.

The upper right panel of \fref{Fig2} depicts the corresponding
magnitude of the ${\bf f}$-vector, as introduced in
\eref{EqFFdagger}. The slightly out-of-phase relation between $f_{0}$
and $f_{1}$ leads to a faster oscillation in $|{\bf f}|$ compared to
the oscillations in $f_{0}$ and $f_{1}$ alone. The very fast
oscillations right of the interface stem from the contributions
$f_{\uparrow \uparrow}$ and $f_{\downarrow \downarrow}$ as shown in
the middle panels.

Finally the lower right panel in \fref{Fig2} depicts the magnitude of
${\bf f}\times{\bf f}^{*}$, giving insight in the spin magnetic
properties of the Cooper pairs. It can be seen that the magnetic
properties of the pairing decay very rapidly on both sides of the
interface, with stronger intensities to be seen in the conical magnet
side. The pairing spin magnetic moments are only visible for
approximately one turn of the conical magnet, while the magnetic
moments in the superconductor side of the interface describe the
so-called inverse proximity effect.

\subsection{Fixed $\alpha$, general $\beta$}
\label{Sec3_2}

\begin{figure}
  \centering \includegraphics[width=\textwidth,clip]{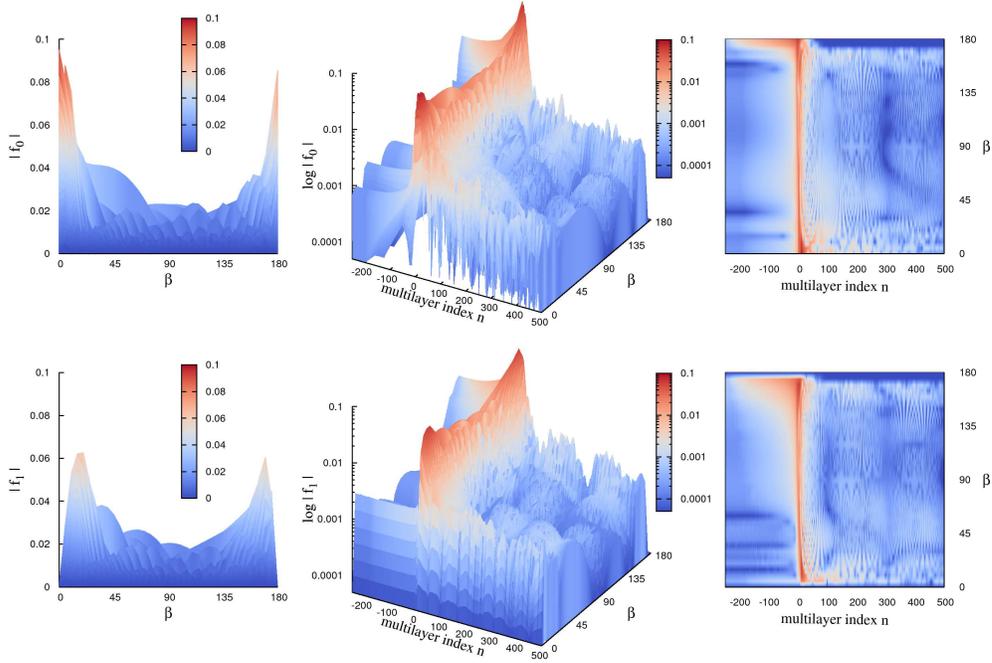}
  \caption{\label{Fig3} Influence of the turning angle $\beta$ on the
    spin-triplet pairing correlations $f_{0}$ (upper panels) and
    $f_{1}$ (lower panels) as defined in
    \eref{EqTripletPairingCorrelations} for fixed opening angle
    $\alpha = 80\,^{\circ}$. Upper (lower) middle panels (logarithmic
    scale) depict full data sets of the magnitudes of $f_{0}$
    ($f_{1}$), whereas in the left (linear scale) and right panels
    (logarithmic scale) the same data is shown when viewed from $yz$
    plane and as top view, respectively.}
\end{figure}

The influence of the turning angle $\beta$ on the spin-triplet pairing
correlations $f_{0}$ and $f_{1}$ is shown in the upper and lower
panels of \fref{Fig3}, respectively.
Keeping the opening angle $\alpha$ fixed to $80\,^{\circ}$ and
allowing the turning angle $\beta$ to vary between $0\,^{\circ}$ and
$180\,^{\circ}$ yields the full data set depicted in the middle panels
of \fref{Fig3} (logarithmic scale). The left (linear scale) and right
panels (logarithmic scale) of \fref{Fig3} show the $yz$ and top view
of the data shown in the middle panels, respectively.

In \fref{Fig3} the maximum values for any $\beta$ clearly stem from
the interface region, with a rapidly decaying oscillation in the
conical magnet region. The influence of $\beta$ on the pairing
correlation $f_{0}$ is largest for $\beta = 0\,^{\circ}$ and $\beta =
180\,^{\circ}$, which is best seen from the upper left panel of
\fref{Fig3}. Going away from those extremal $\beta$ values, the
pairing correlation decays very quickly to approximately one third of
its maximum value for $\beta = 90\,^{\circ}$.

Looking at the $f_{1}$ pairing correlations in the lower panels of
\fref{Fig3} the maxima are slightly shifted away from the extremal
$\beta$ values but show again a decrease towards $\beta =
90\,^{\circ}$ and fall again down to roughly one third of the maximum
$f_{1}$. In contrast to $f_{0}$ in the upper panels with maximal
contributions from $\beta = 0\,^{\circ}$ and $\beta = 180\,^{\circ}$,
the respective $\beta$ contributions to $f_{1}$ vanish. The right
panels show a top view of the data shown in the middle panels to
depict the slight oscillations with changing $\beta$. These patterns
are similar for $f_{0}$ and $f_{1}$.

Keeping in mind that the data displayed in \fref{Fig3} corresponds to
$\alpha = 80\,^{\circ}$, the middle and right panels allow an estimate
of the turning angle's influence on the decay length of $f_{0}$ and
$f_{1}$. Starting from the nonvanishing contributions to $f_{0}$ for
$\beta = 0\,^{\circ}$ ($\beta = 180\,^{\circ}$) in the interface
region, the decay length is largest (smallest) for the data shown
here. For the other values of $\beta$ the decay length has the same
order of magnitude, being influenced by some oscillations due to the
multilayer index $n$. Looking now at the $f_{1}$ values close to
$\beta = 0\,^{\circ}$ ($\beta = 180\,^{\circ}$) in the interface
region the decay length shows a similar behaviour to $f_{0}$, being
largest (smallest) for $\beta = 0\,^{\circ}$ ($\beta = 180\,^{\circ}$)
values. Again, for the other values of $\beta$ we observe decay
lengths of the same order of magnitude and similar to the ones seen
for $f_{0}$.

\begin{figure}
  \centering \includegraphics[width=\textwidth,clip]{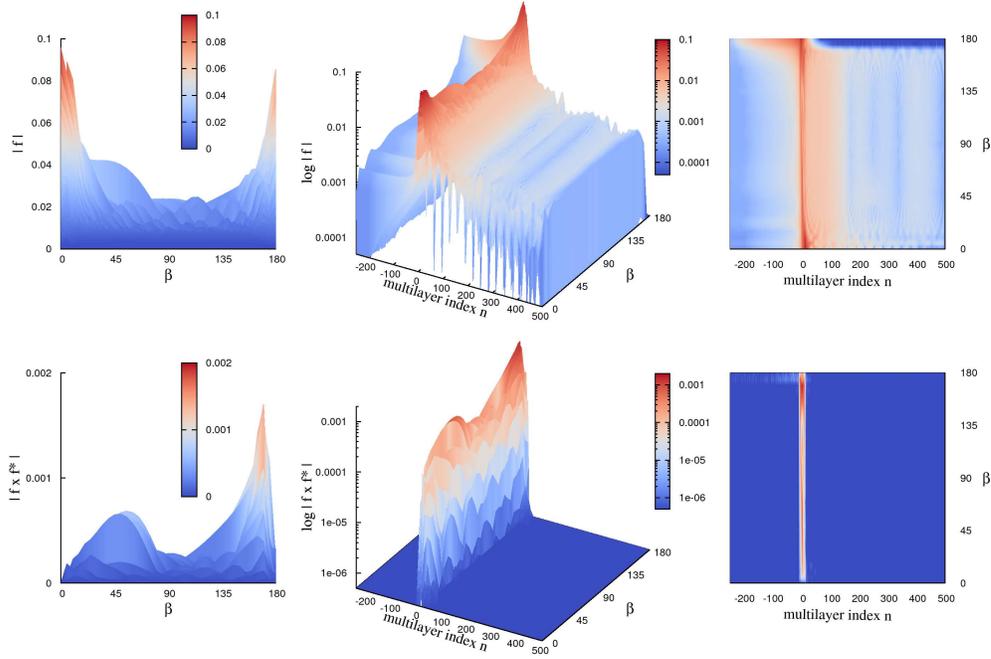}
  \caption{\label{Fig4} Influence of the turning angle $\beta$ on the
    magnitude of the ${\bf f}$-vector and ${\bf f}\times{\bf f}^{*}$
    as defined in \eref{EqFFdagger} for fixed opening angle $\alpha =
    80\,^{\circ}$. Logarithmically scaled upper (lower) middle panels
    depict full data sets of the magnitudes of the ${\bf f}$-vector
    (${\bf f}\times{\bf f}^{*}$), whereas in the left (linear scale)
    and right panels (logarithmic scale) the same data is shown when
    viewed from $yz$ plane and as top view, respectively.}
\end{figure}

\Fref{Fig4} now shows the influence of $\beta$ on the magnitudes of
the ${\bf f}$-vector and ${\bf f}\times{\bf f}^{*}$, as introduced in
\eref{EqFFdagger}, respectively. The upper panels present the
magnitude of ${\bf f}$ and the lower panels the magnitude of ${\bf
  f}\times{\bf f}^{*}$, each shown in the same parameter ranges and
views as \fref{Fig3}.
Not surprisingly, the magnitude of the ${\bf f}$-vector resembles the
appearance of $f_{0}$ and $f_{1}$ from \fref{Fig3}. The maxima occur
for $\beta = 0\,^{\circ}$ and $\beta = 180\,^{\circ}$, as for $f_{0}$,
whereas the decaying patterns are similar to the ones already familiar
from \fref{Fig3}. Additionally, the most prominent contributions to
the triplet pairing correlations again originate from the interface
region. Only in the top view figure (upper right panel) can be seen a
somewhat blurred superposition of the right panels of
\fref{Fig3}. This can be attributed to the slight out-of-phase
relation of $f_{0}$ and $f_{1}$, as already seen in \fref{Fig2} and
discussed in \sref{Sec3_1}.

However, the magnitude of ${\bf f}\times{\bf f}^{*}$ describing the
triplet spin magnetic properties behaves surprisingly differently. As
seen in the lower panels of \fref{Fig4}, it is zero for $\beta =
0\,^{\circ}$, then increases up to a first maximum around $\beta =
45\,^{\circ}$, and subsequently decreases again to form a small
plateau around $\beta = 90\,^{\circ}$. After that the magnitude of
${\bf f}\times{\bf f}^{*}$ has another sharp increase to its maximum
value, before vanishing again for $\beta = 180\,^{\circ}$.  The fact
that this vanishes for $\beta = 0\,^{\circ}$ and $\beta =
180\,^{\circ}$ is to be expected since these are collinear magnetic
structures (ferro- or antiferromagnetic, respectively) and so there
are no spin-flip processes at the SC/CM interface.

\subsection{General $\alpha$, general $\beta$}
\label{Sec3_3}

\begin{figure}
  \centering \includegraphics[width=0.9\textwidth,clip]{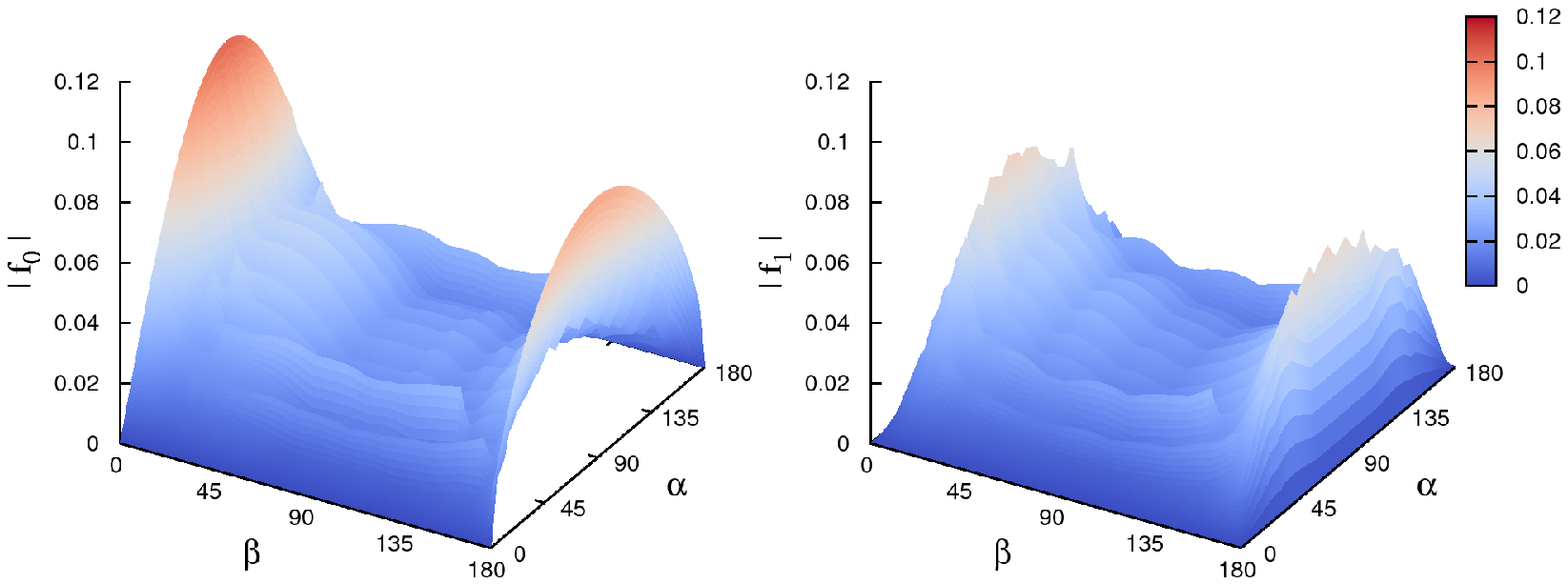}
  \caption{\label{Fig5} Left (right) panel: $yz$ views of the
    magnitudes of $f_{0}$ ($f_{1}$) as depicted in the upper (lower)
    left panel of \fref{Fig4} for varying opening angle $\alpha$.}
\end{figure}

We now consider variations in the full range of both conical magnet
angles $\alpha$ and varying $\beta$. As has been discussed in
\sref{Sec3_2} the most prominent contributions to either $f_{0}$ and
$f_{1}$ or $|{\bf f}|$ and $|{\bf f}\times{\bf f}^{*}|$ originate from
the interface region, as can be seen clearly from \fref{Fig3} and
\fref{Fig4}. The height profile of the respective quantities for a
fixed $\alpha$ and varying $\beta$ are shown in the left panels of
those figures, displaying the $yz$ views. Since the maxima of $f_{0}$
and $f_{1}$ are always occurring in the interface region this gives us
the opportunity to have a look at the data for all angles $\alpha$ and
$\beta$. This is displayed in the plots shown in \fref{Fig5} with the
left and right panels depicting the $yz$ views of $f_{0}$ and $f_{1}$
now shown with varying opening angle $\alpha$.
As can be seen, for $\alpha = 0\,^{\circ}$ and $\alpha =
180\,^{\circ}$ the contributions to $f_{0}$ and $f_{1}$ vanish. With
increasing $\alpha$ the triplet pairing correlations $f_{0}$ and
$f_{1}$ increase up to a maximum around $\alpha = 90\,^{\circ}$ to
then decay to zero again afterwards. In \fref{Fig5} the appearance
along the $\beta$ coordinate is similar for all $\alpha$ values and
has already been shown, as for example in \fref{Fig3}.

\begin{figure}
  \centering \includegraphics[width=0.9\textwidth,clip]{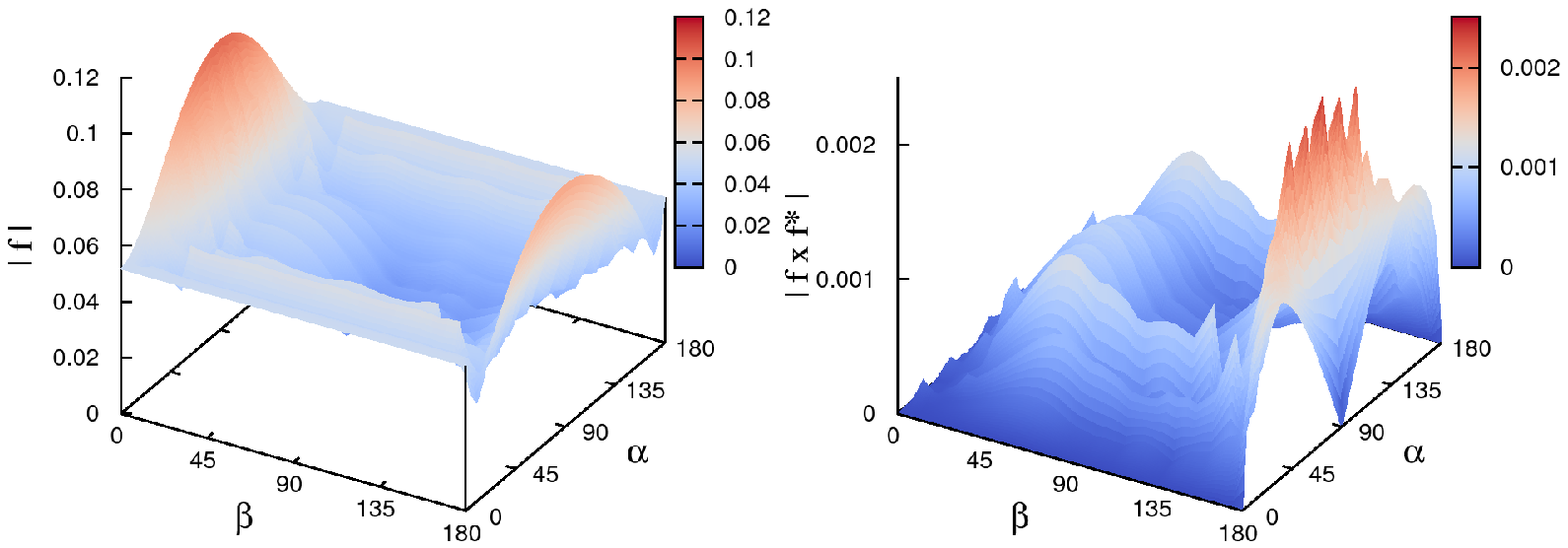}
  \caption{\label{Fig6} Left (right) panel: $yz$ views of the
    magnitudes of the ${\bf f}$-vector and ${\bf f}\times{\bf f}^{*}$
    as depicted in the upper (lower) left panel of \fref{Fig5} for
    varying opening angle $\alpha$.}
\end{figure}

The full angle dependence on the magnitudes of the ${\bf f}$-vector
and ${\bf f}\times{\bf f}^{*}$ is shown in the left and right panels
of \fref{Fig6}, respectively.
Focusing for the moment on the left panel showing the $yz$ views of
the magnitude of the ${\bf f}$-vector one notices that for $\alpha =
0\,^{\circ}$ and $\alpha = 180\,^{\circ}$ it has a nonvanishing
contribution, in contrast to the vanishing contributions of $f_{0}$
and $f_{1}$ depicted in \fref{Fig5}. This can be understood knowing
that vanishing contributions to $f_{1}$ are due to equal size but
opposite sign contributions from $f_{\uparrow \uparrow}$ and
$f_{\downarrow \downarrow}$ generating $f_{1}$. Looking additionally
at the separate contributions $f_{\uparrow \uparrow}$ and
$f_{\downarrow \downarrow}$ can provide a deeper understanding of
spin-triplet generation at specific interfaces. Coming back to the
data at hand, maxima in the magnitude of the ${\bf f}$-vector appear
for $\alpha = 90\,^{\circ}$ and $\beta$ either $0\,^{\circ}$ or
$180\,^{\circ}$, i.e., a complete ferromagnet or antiferromagnet with
the magnetic moments oriented along the $z$-axis. All deviations from
this ferromagnetic-like arrangement of magnetic moments between
adjacent layers, either changing just $\beta$ (corresponding to a
helical magnet), or changing both $\alpha$ and $\beta$ (a CM) leads to
a decrease in the magnitude of the ${\bf f}$-vector associated with
the unitary triplet pairing correlations in the SC/CM interface.

The right panel of \fref{Fig6} now shows the magnitude of ${\bf
  f}\times{\bf f}^{*}$ associated with the non-unitary triplet
pairing, or spin magnetic moment of the Cooper pair. The pattern of
this quantity with respect to varying $\beta$ are the same as already
discussed in \sref{Sec3_2} and shown in \fref{Fig4}, namely vanishing
contributions for $\beta = 0\,^{\circ}$, increase up to $\beta =
45\,^{\circ}$, again decreasing to the plateau around $\beta =
90\,^{\circ}$, and then the sharp increase to the maximum value. The
maximum values of the non-unitary pairing correlations clearly occur
for values of $\alpha$ near to $90\,^{\circ}$ and $\beta =
180\,^{\circ}$, although the correlation becomes zero exactly at the
point $\alpha= 90\,^{\circ}$, $\beta = 180\,^{\circ}$, since this is a
antiferromagnetic structure.

One can see in these plots that the CM angle values of $\alpha =
80\,^{\circ}$ and $\beta = 30\,^{\circ}$, corresponding to Ho as used
in the experiments, are fairly typical in their behaviour. In fact in
\fref{Fig6} there is a broad maximum in the non-unitary triplet
pairing function ${\bf f}\times{\bf f}^{*}$ for values similar to
those of Ho.

\section{Summary and Outlook}
\label{SummaryAndOutlook}

In summary, we have presented results on the spin-triplet pairing
correlations in heterostructures built up from an $s$-wave
superconductor and a general conical magnet described by an opening
angle $\alpha$ and a turning (or pitch) angle $\beta$. All the results
shown are obtained by self-consistent solutions of the microscopic
Bogoliubov-de Gennes equations in the clean limit. The influence of
the opening angle $\alpha$ and the turning (or pitch) angle $\beta$ on
the induced spin-triplet pairing correlations has been investigated in
detail. We see that both unitary and non-unitary triplet pairing
components are generated by any non-collinear magnetic structure, and
that the CM values for Ho are typical of the behaviour over a very
wide range of variations in these angles.

\ack This work has been financially supported by the EPSRC
(EP/I037598/1) and made use of computational resources of the
University of Bristol. The authors gratefully acknowledge discussions
with M. G. Blamire and J. W. A. Robinson on Ho containing
heterostructures.

B. Gy\"orffy was an inspiration in the original plan for this project,
and he was a co-investigator on this EPSRC grant. Sadly he passed away
before the first results of this project were obtained.

\end{document}